\documentclass[conference]{IEEEtran}
\IEEEoverridecommandlockouts

\usepackage{cite}

\ifCLASSINFOpdf
  \usepackage[pdftex]{graphicx}
  \usepackage{epstopdf} 
\else
\fi
\usepackage{amsmath}
\usepackage{algorithm}
\usepackage{algorithmic}

  \usepackage[caption=false,font=footnotesize]{subfig}
\usepackage{url}

\usepackage{amsthm}

\newtheoremstyle{mydefinition}
{}
{}
{}
{0pt}
{\bfseries}
{.}
{ }
{\thmname{#1}\thmnumber{ #2}: \thmnote{#3}}

\theoremstyle{mydefinition}

\newtheoremstyle{myremark}
{}
{}
{}
{0pt}
{\bfseries}
{.}
{ }
{\thmname{#1}\thmnumber{ #2}: \thmnote{#3}}
\theoremstyle{myremark}

\newtheoremstyle{remarkshort}
{}
{}
{}
{0pt}
{\bfseries}
{.}
{ }
{\thmname{#1}\thmnumber{ #2}}

\theoremstyle{remarkshort}

\theoremstyle{remarkshort}

\usepackage{amssymb,amsmath}
\usepackage{xspace}
\usepackage{bm}
\usepackage{bbm}
\usepackage{mathtools}

\let\originalleft\left
\let\originalright\right
\renewcommand{\left}{\mathopen{}\mathclose\bgroup\originalleft}
\renewcommand{\right}{\aftergroup\egroup\originalright}

\newcommand{\set}[1]{\mathcal{#1}} 
\newcommand{\mat}[1]{\mathbf{#1}} 
\renewcommand{\vec}[1]{\mathbf{#1}} 

\newcommand{\parens}[1]{{\left(#1\right)}\xspace}

\newcommand{\braces}[1]{{\left\{#1\right\}}\xspace}
\newcommand{\bars}[1]{{\left\vert#1\right\vert}\xspace}

\newcommand{\complex}{{\mathbb{C}}\xspace}

\newcommand{\logtwo}[1]{\ensuremath{\mathrm{log}_{2}\parens{#1}}}

\DeclarePairedDelimiter{\nint}\lfloor\rceil

\newcommand{\card}[1]{\bars{#1}}

\newcommand{\setcomplex}{{\complex}}

\newcommand{\setvector}[2]{#1^{#2 \times 1}}

\newcommand{\setvectorcomplex}[1]{\setvector{\setcomplex}{#1}}

\newcommand{\setmatrix}[3]{#1^{#2 \times #3}}

\newcommand{\setmatrixcomplex}[2]{\setmatrix{\setcomplex}{#1}{#2}}

\newcommand{\ctrans}{^{{*}}}

\DeclareMathOperator*{\argmin}{\mathrm{argmin}}

\newcommand{\subjectto}{\mathrm{s.t.~}}
\newcommand{\subto}{\subjectto}

\newcommand{\powertx}{P_{\mathrm{tx}}}

\newcommand{\powernoise}{P_{\mathrm{n}}}

\newcommand{\snr}{{\mathsf{SNR}}}

\newcommand{\sinr}{{\mathsf{SINR}}}

\newcommand{\inr}{{\mathsf{INR}}}

\newcommand{\Nt}{N_\mathrm{t}} 
\newcommand{\Nr}{N_\mathrm{r}} 
\newcommand{\labeltx}{\mathrm{tx}}
\newcommand{\labelrx}{\mathrm{rx}}

\def\va{{\vec{a}}}

\def\vf{{\vec{f}}}

\def\vh{{\vec{h}}}

\def\vw{{\vec{w}}}

\def\vbeta{\bm{\beta}}

\def\mH{{\mat{H}}}

\def\mOmega{{\mat{\Omega}}}

\def\sF{{\set{F}}}
\def\sG{{\set{G}}}

\def\sN{{\set{N}}}

\def\sT{{\set{T}}}

\def\sW{{\set{W}}}

\newcommand{\vatx}{\va_{\labeltx}\xspace}
\newcommand{\varx}{\va_{\labelrx}\xspace}

\newcommand{\labelul}{\mathsf{UL}}  
\newcommand{\thetaul}{\theta^{\labelul}}
\newcommand{\phiul}{\phi^{\labelul}}

\newcommand{\varthetaul}{\vartheta^{\labelul}}
\newcommand{\varphiul}{\varphi^{\labelul}}
\newcommand{\snrul}{\snr^{\labelul}}   
\newcommand{\powerultx}{\powertx^{\labelul}}    
\newcommand{\powerulnoise}{\powernoise^{\labelul}}  
\newcommand{\gainulrx}{G^{\labelul}_{\labelrx}}    
\newcommand{\vhul}{\vh^{\labelul}}
\newcommand{\vhult}{\vhul_t}

\newcommand{\labeldl}{\mathsf{DL}}
\newcommand{\thetadl}{\theta^{\labeldl}}
\newcommand{\phidl}{\phi^{\labeldl}}

\newcommand{\varthetadl}{\vartheta^{\labeldl}}
\newcommand{\varphidl}{\varphi^{\labeldl}}
\newcommand{\snrdl}{\snr^{\labeldl}}   
\newcommand{\inrdl}{\inr^{\labeldl}}   
\newcommand{\sinrdl}{\sinr^{\labeldl}}   
\newcommand{\powerdltx}{\powertx^{\labeldl}}   
\newcommand{\powerdlnoise}{\powernoise^{\labeldl}}   
\newcommand{\gaindltx}{G^{\labeldl}_{\labeltx}}    
\newcommand{\vhdl}{\vh^{\labeldl}}
\newcommand{\vhdlt}{\vhdl_t}

\newcommand{\thphult}{\parens{\thetaul_t,\phiul_t}}
\newcommand{\thphdlt}{\parens{\thetadl_t,\phidl_t}}

\usepackage{xspace}
\usepackage[acronym,nogroupskip,nonumberlist,nopostdot]{glossaries}
\makeglossaries

\newcommand{\steer}{\textsc{Steer}\xspace}

\usepackage{xcolor}

\newacronym{snr}{SNR}{signal-to-noise ratio}
\newacronym{sinr}{SINR}{signal-to-interference-plus-noise ratio}
\newacronym{inr}{INR}{interference-to-noise ratio}
\newacronym{sir}{SIR}{signal-to-interference ratio}
\newacronym{sqr}{SQR}{signal-to-quantization-noise ratio}
\newacronym{sqnr}{SQNR}{signal-to-quantization-plus-noise ratio}
\newacronym{ian}{IAN}{interference as noise}
\newacronym{ber}{BER}{bit error rate}
\newacronym{pn}{PN}{pseudorandom noise}
\newacronym{bfsk}{BFSK}{binary frequency shift keying}
\newacronym{fh}{FH}{frequency-hopped}
\newacronym{fh-bfsk}{FH-BFSK}{frequency-hopped binary frequency shift keying}
\newacronym{crc}{CRC}{cyclic redundancy check}
\newacronym{isi}{ISI}{intersymbol interference}
\newacronym{dsss}{DSSS}{direct-sequence spread spectrum}
\newacronym{ofdm}{OFDM}{orthogonal frequency-division multiplexing}
\newacronym{ofdma}{OFDMA}{orthogonal frequency-division multiple access}
\newacronym{sdr}{SDR}{software-defined radio}
\newacronym{tx}{TX}{transmitter}
\newacronym{rx}{RX}{receiver}
\newacronym{fdd}{FDD}{frequency-division duplexing}
\newacronym{tdd}{TDD}{time-division duplexing}
\newacronym{fdma}{FDMA}{frequency-division multiple access}
\newacronym{tdma}{TDMA}{time-division multiple access}
\newacronym{sdma}{SDMA}{space-division multiple access}
\newacronym[plural=MPCs]{mpc}{MPC}{multipath component}
\newacronym{mui}{MUI}{multi-user interference}
\newacronym{lsb}{LSB}{least significant bit}
\newacronym{jcas}{JCAS}{joint communication and sensing}

\newacronym{qam}{QAM}{quadrature amplitude modulation}
\newacronym{mqam}{MQAM}{M-ary quadrature amplitude modulation}

\newacronym{ls}{LS}{least-squares}
\newacronym{lms}{LMS}{least mean squares}
\newacronym{rls}{RLS}{recursive least-squares}
\newacronym{rzf}{RZF}{regularized zero-forcing}
\newacronym{mmse}{MMSE}{minimum mean square error}
\newacronym{lmmse}{LMMSE}{linear minimum mean square error}
\newacronym{mse}{MSE}{mean square error}
\newacronym{fft}{FFT}{fast Fourier transform}
\newacronym{dft}{DFT}{discrete Fourier transform}
\newacronym{dtft}{DTFT}{discrete-time Fourier transform}
\newacronym{ctft}{CTFT}{continuous-time Fourier transform}
\newacronym{ml}{ML}{machine learning}
\newacronym[plural=NNs]{nn}{NN}{neural network}
\newacronym[plural=RNNs]{rnn}{RNN}{recurrent neural network}
\newacronym[plural=ADCs]{adc}{ADC}{analog-to-digital converter}
\newacronym[plural=DACs]{dac}{DAC}{digital-to-analog converter}
\newacronym[plural=FPGAs]{fpga}{FPGA}{field-programmable gate array}
\newacronym{evm}{EVM}{error vector magnitude}
\newacronym{enob}{ENOB}{effective number of bits}
\newacronym{zf}{ZF}{zero-forcing}
\newacronym{rv}{r.v.}{random variable}
\newacronym{omp}{OMP}{orthogonal matching pursuit}
\newacronym{svd}{SVD}{singular value decomposition}

\newacronym{sdp}{SDP}{semidefinite programming}
\newacronym{psd}{PSD}{positive semidefinite}
\newacronym{nsd}{NSD}{negative semidefinite}

\newacronym{ks}{K-S}{Kolmogorov-Smirnov}

\newacronym{mad}{MAD}{median absolute deviation around the median}

\newacronym{agc}{AGC}{automatic gain control}
\newacronym{rf}{RF}{radio frequency}
\newacronym{if}{IF}{intermediate frequency}
\newacronym{los}{LOS}{line-of-sight}
\newacronym{nlos}{NLOS}{non-line-of-sight}
\newacronym{ple}{PLE}{path loss exponent}
\newacronym[plural=dB,firstplural=decibels (dB)]{db}{dB}{decibel}
\newacronym[plural=dBm,firstplural=decibel milliwatts (dBm)]{dbm}{dBm}{decibel milliwatts}
\newacronym{pa}{PA}{power amplifier}
\newacronym{lna}{LNA}{low noise amplifier}
\newacronym{vga}{VGA}{variable gain amplifier}
\newacronym{cw}{CW}{continuous wave}
\newacronym{papr}{PAPR}{peak-to-average power ratio}
\newacronym{usrp}{USRP}{Universal Software Radio Peripheral}
\newacronym{irr}{IRR}{image rejection ratio}
\newacronym{lo}{LO}{local oscillator}
\newacronym{vm}{VM}{vector modulator}
\newacronym{mmwave}{mmWave}{millimeter wave}
\newacronym{eirp}{EIRP}{effective isotropic radiated power}
\newacronym{rsrp}{RSRP}{reference signal received power}

\newacronym{csma}{CSMA}{carrier-sense multiple access}
\newacronym{csmaca}{CSMA/CA}{carrier-sense multiple access with collision avoidance}
\newacronym{csmacd}{CSMA/CD}{carrier-sense multiple access with collision detection}
\newacronym{mac}{MAC}{medium access control}
\newacronym{phy}{PHY}{physical layer}
\newacronym{4g}{4G}{fourth generation}
\newacronym{lte}{LTE}{Long-Term Evolution}
\newacronym{4glte}{4G LTE}{\gls{4g} \gls{lte}}
\newacronym{5g}{5G}{fifth generation}
\newacronym{nr}{NR}{New Radio}
\newacronym{5gnr}{5G NR}{5G New Radio}
\newacronym{ieee}{IEEE}{Institute of Electrical and Electronics Engineers}
\newacronym{wifi}{Wi-Fi}{IEEE 802.11}
\newacronym{lan}{LAN}{local area network}
\newacronym{wlan}{WLAN}{wireless local area network}
\newacronym[plural=BSs]{bs}{BS}{base station}
\newacronym[plural=SBSs]{sbs}{SBS}{small-cell base station}
\newacronym[plural=FD-SBSs]{fdsbs}{FD-SBS}{\gls{fd}-enabled \gls{sbs}}
\newacronym[plural=MBSs]{mbs}{MBS}{macrocell base station}
\newacronym[plural=UEs]{ue}{UE}{user equipment}
\newacronym{ul}{UL}{uplink}
\newacronym{dl}{DL}{downlink}
\newacronym{qos}{QoS}{Quality of Service}
\newacronym{fcc}{FCC}{Federal Communications Commission}
\newacronym{iab}{IAB}{integrated access and backhaul}
\newacronym{fab}{FAB}{fixed access and backhaul}
\newacronym{hetnet}{HetNet}{heterogeneous network}

\newacronym{siso}{SISO}{single-input single-output}
\newacronym{mimo}{MIMO}{multiple-input multiple-output}
\newacronym{sumimo}{SU-MIMO}{single-user \gls{mimo}}
\newacronym{mumimo}{MU-MIMO}{multi-user \gls{mimo}}
\newacronym{bf}{BF}{beamforming}
\newacronym{ca}{CA}{constant amplitude}
\newacronym{ula}{ULA}{uniform linear array}
\newacronym{upa}{UPA}{uniform planar array}
\newacronym[\glslongpluralkey={angles of arrival}]{aoa}{AoA}{angle of arrival}
\newacronym[\glslongpluralkey={angles of departure}]{aod}{AoD}{angle of departure}
\newacronym{dof}{DoF}{degrees of freedom}
\newacronym{csi}{CSI}{channel state information}
\newacronym{csit}{CSIT}{\gls{csi} at the transmitter}
\newacronym{csir}{CSIR}{\gls{csi} at the receiver}
\newacronym{cs}{CS}{compressed sensing}

\newacronym{fd}{FD}{in-band full-duplex}
\newacronym{hd}{HD}{half-duplex}
\newacronym{si}{SI}{self-interference}
\newacronym{sic}{SIC}{self-interference cancellation}
\newacronym{soi}{SoI}{signal of interest}
\newacronym{asic}{A-SIC}{analog \acrlong{sic}}
\newacronym{dsic}{D-SIC}{digital \gls{sic}}
\newacronym{star}{STAR}{simultaneous transmit and receive}
\newacronym{warp}{WARP}{Wireless Open-Access Research Platform}
\newacronym{bfc}{BFC}{beamforming cancellation}
\newacronym{ipi}{IPI}{inter-panel-interference}
\newacronym{ipic}{IPIC}{inter-panel-interference cancellation}

\newacronym{qcqp}{QCQP}{quadratically-constrained quadratic programming}
\newacronym{pdf}{PDF}{probability density function}
\newacronym{cdf}{CDF}{cumulative distribution function}
\newacronym{iid}{i.i.d.}{independently and identically distributed}

\newacronym{elf}{ELF}{extremely low frequency}
\newacronym{slf}{SLF}{super low frequency}
\newacronym{ulf}{ULF}{ultra low frequency}
\newacronym{vlf}{VLF}{very low frequency}
\newacronym{lf}{LF}{low frequency}
\newacronym{mf}{MF}{medium frequency}
\newacronym{hf}{HF}{high frequency}
\newacronym{vhf}{VHF}{very high frequency}
\newacronym{uhf}{UHF}{ultra high frequency}
\newacronym{shf}{SHF}{super high frequency}
\newacronym{ehf}{EHF}{extremely high frequency}
\newacronym{thf}{THF}{tremendously high frequency}

\newacronym{wncg}{WNCG}{Wireless Networking and Communications Group}
\newacronym{linc}{LINC}{Laboratory of Informatics, Networks, and Communications}
\newacronym{ut}{UT Austin}{The University of Texas at Austin}
\newacronym{uiuc}{UIUC}{University of Illinois at Urbana-Champaign}
\newacronym{usc}{USC}{University of Southern California}
\newacronym{mit}{MIT}{Massachusetts Institute of Technology}
\newacronym{berkeley}{UC Berkeley}{University of California, Berkeley}
\newacronym{osu}{OSU}{Ohio State University}

\newacronym{leo}{LEO}{low Earth orbit}

\newcommand{\leo}{\gls{leo}\xspace}

\newcommand{\mmwave}{\gls{mmwave}\xspace}

\newcommand{\gcdf}{\gls{cdf}\xspace}
\newcommand{\gpcdf}{\glspl{cdf}\xspace}

\newcommand{\gsnr}{\gls{snr}\xspace}

\newcommand{\gsinr}{\gls{sinr}\xspace}

\newcommand{\gpsnr}{\glspl{snr}\xspace}

\newcommand{\gpsinr}{\glspl{sinr}\xspace}

\newcommand{\figref}[1]{\figurename~\ref{#1}}
\newcommand{\algref}[1]{Algorithm~\ref{#1}}

\usepackage{mathtools}
\usepackage{comment}

\definecolor{uclablue}{RGB}{39,116,174}
\definecolor{uclabluedarkest}{RGB}{0,59,92}
\definecolor{uclabluedarker}{RGB}{0,85,135}
\definecolor{uclabluelighter}{RGB}{139,184,232}
\definecolor{uclabluelightest}{RGB}{218,235,254}

\definecolor{uclagold}{RGB}{255,209,0}
\definecolor{uclagolddarker}{RGB}{255,199,44}
\definecolor{uclagolddarkest}{RGB}{255,184,28}

\definecolor{uclamagenta}{RGB}{255,0,165}

\usepackage{enumitem}

\usepackage{pgfplots}
\pgfplotsset{compat=newest}
\usetikzlibrary{plotmarks}
\usetikzlibrary{arrows.meta}
\usepgfplotslibrary{patchplots}
\usepackage{grffile}

\pgfplotsset{plot coordinates/math parser=false}

\makeatletter
\newcommand\fs@betterruled{%
    \def\@fs@cfont{\bfseries}\let\@fs@capt\floatc@ruled
    \def\@fs@pre{\vspace*{5.5pt}\hrule height.8pt depth0pt \kern2pt}%
    \def\@fs@post{\kern2pt\hrule\relax}%
    \def\@fs@mid{\kern2pt\hrule\kern2pt}%
    \let\@fs@iftopcapt\iftrue}
\floatstyle{betterruled}
\restylefloat{algorithm}
\makeatother

\begin{document}

%
\title{Beam Tracking for Full-Duplex User Terminals in\\Low Earth Orbit Satellite Communication Systems}

%
%
\author{
\IEEEauthorblockN{Chaeyeon Kim\IEEEauthorrefmark{1}, Joohyun Son\IEEEauthorrefmark{1}\IEEEauthorrefmark{2}, Daesik Hong\IEEEauthorrefmark{2}, and Ian P.~Roberts\IEEEauthorrefmark{1}}%
\IEEEauthorblockA{\IEEEauthorrefmark{1}Wireless Lab, Department of Electrical and Computer Engineering, University of California, Los Angeles, CA USA}%
\IEEEauthorblockA{\IEEEauthorrefmark{2}Information and Telecommunication Lab, School of Electrical and Electronic Engineering, Yonsei University, Seoul, Korea}%
}

\maketitle

\begin{abstract}
This paper introduces a novel beam tracking scheme for full-duplex ground user terminals aiming to transmit uplink and receive downlink from two low Earth orbit (LEO) satellites at the same time and same frequency.
Our proposed technique leverages observed phenomena from a recent measurement campaign to strategically select transmit and receive beams which couple low self-interference across the satellites' trajectories, thereby enabling in-band full-duplex operation. 
Our scheme takes a measurement-driven approach, meaning it does not rely on explicit knowledge of the self-interference channel and can inherently account for hardware impairments or other nonidealities.
We show that our proposed scheme reliably selects beams which spatially cancel self-interference to below the noise floor, circumventing the need for digital/analog cancellation.
Simulation results using satellite and orbital parameters published in 3GPP and FCC filings show that this substantial reduction in self-interference does not prohibitively compromise beamforming gain, allowing the user terminal to attain near-maximal SINRs, thus unlocking full-duplex operation. 
\end{abstract}

\glsresetall

\section{Introduction} \label{sec:introduction}

\Gls{leo} satellite communication systems continue to cement their role in delivering near-global connectivity in the 6G era \cite{IMT_framework}.
Large-scale efforts by SpaceX, OneWeb, and Amazon aim to deploy tens of thousands of \leo satellites through the end of the decade and beyond to provide un-/under-served ground users across the globe with broadband wireless connectivity.
While undoubtedly revolutionary, dense \leo networks face noteworthy challenges involving user data rates, latency, handover, coexistence, and duplexing flexibility \cite{kodheli_satellite}.
In pursuit of overcoming these challenges, this paper aims its sights on upgrading ground user terminals with in-band full-duplex capability: the ability to transmit and receive at the same time and same frequency \cite{smida_fd_phy_2024}.

Like other wireless networks, in-band full-duplex capability stands to be a transformational physical layer upgrade in \leo satellite communication systems \cite{baeza_overview_2023}.
While there are certainly ripe applications of full-duplex at the satellite side \cite{baeza_overview_2023}, this work focuses specifically on full-duplex ground user terminals.
Full-duplex capability would potentially double data rates by allowing users to transmit uplink while receiving downlink at the same frequency. 
Further, spectrum sensing and cognitive radio could be harnessed to enable new spectrum sharing paradigms across \leo constellations.
Beyond this, more flexible duplexing, such as \acrlong{tdd}, could be made possible in \leo systems without prohibitively long guard intervals to accommodate the timing advances necessary to overcome millisecond-long propagation delays \cite{Flexible}.

Extensive research over the past decade has developed techniques to realize in-band full-duplex capability, with most of this focus on conventional terrestrial wireless systems like 5G/6G cellular and Wi-Fi networks \cite{smida_fd_phy_2024}.
While analog and digital self-interference cancellation techniques have proven effective in lower-frequency radios \cite{smida_fd_phy_2024}, they are not particularly attractive in high-frequency transceivers like those used in \leo \cite{roberts_wcm}.
The high frequencies, wide bandwidths, and antenna arrays used in \leo user terminals complicate the extension of existing analog and digital self-interference cancellation solutions to enable full-duplex \cite{roberts_wcm}.
Other full-duplex solutions \cite{choi_circular_2024, erkelenz_hybrid_2024, sazegar_kuband_2023, navarro_ultra_2017, abedian_mmwave_2023} have leveraged \acrlong{rf} components and electromagnetic isolation to develop custom user terminals and antennas, but the majority \cite{choi_circular_2024, erkelenz_hybrid_2024,sazegar_kuband_2023, navarro_ultra_2017} have focused on sub-band full-duplex rather than in-band full-duplex.

More recently, full-duplex solutions which harness beamforming to cancel self-interference spatially have been developed for \mmwave radios similar to \leo user terminals and have proven capable of suppressing self-interference to near or even below the noise floor without the need for supplemental cancellation \cite{roberts_wcm}. 
Among these beamforming solutions, however, often encountered are two noteworthy practical shortcomings.
First, many ignore hardware constraints associated with analog beamforming systems, such as finite-resolution phase shifters and array miscalibration. 
Second, many solutions rely on explicit knowledge of the self-interference channel, though such knowledge is difficult to acquire accurately \cite{roberts_wcm}. 
These two shortcomings bring into question the practicality of such techniques in \leo ground user terminals.
One notable existing solution called \steer \cite{roberts_steer} overcomes the two aforementioned challenges by taking a measurement-driven approach that exploits the variability of self-interference over small spatial neighborhoods \cite{roberts_att_angular}. 

Similar to \steer \cite{roberts_steer}, we propose harnessing this small-scale spatial variability to construct a beam tracking technique to enable full-duplex ground user terminals in \leo satellite networks. 
As a first step, our proposed solution takes advantage of the known orbitals of satellites to populate a set of candidate beams by constructing a grid of beams along the trajectories of uplink and downlink satellites of interest. 
Then, through a series of measurements taken solely at the user terminal, our solution chooses the beam pair which maximizes the sum spectral efficiency. 
Simulation results indicate that, with our technique, a user terminal can reliably reduce self-interference to below the noise floor while delivering high gain toward the uplink and downlink satellites.
This allows it to attain near-maximal \gls{sinr} on the downlink and near-maximal \gls{snr} on the uplink, unlocking full-duplex operation. 

\begin{figure}
    \centering
    \includegraphics[width=\linewidth,height=\textheight,keepaspectratio]{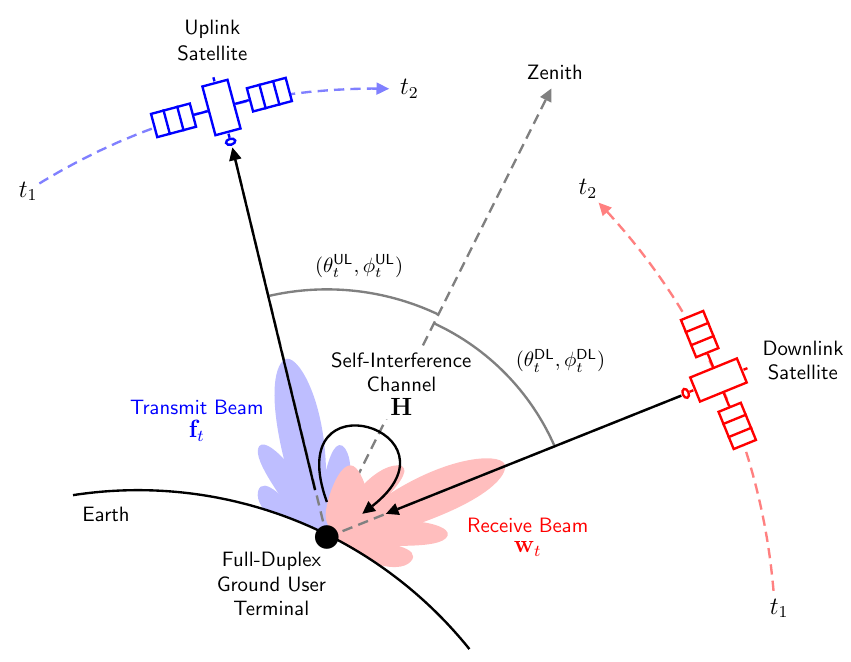}
    \caption{At time $t$, a full-duplex ground user terminal transmits uplink to a satellite while receiving downlink from another satellite at the same frequency. In doing so, self-interference is coupled by its transmit and receive beams across the channel $\mH$, the severity of which is proportional to $\bars{\vw_t\ctrans \mH \vf_t}^2$. This work aims to design $\vf_t$ and $\vw_t$ for all $t$ to reduce self-interference while delivering high uplink and downlink SNR.}
    \label{fig:system}
\end{figure}

\section{System Model} \label{sec:system-model}

This work considers a single ground user terminal being served by a LEO satellite constellation.
More specifically, as shown in \figref{fig:system}, we consider a user terminal interested in receiving downlink from one satellite and transmitting uplink from another satellite at the same time and same frequency. 
The goal of this work is to enable such in-band full-duplex operation at the user terminal.

Suppose the user terminal is equipped with two separate antenna arrays, one for transmission and one for reception, with each array controlled independent of the other, as is common in traditional half-duplex user terminals.
For simplicity, assume the user terminal's transmit and receive arrays are half-wavelength uniform planar arrays with $\Nt$ and $\Nr$ antenna elements, respectively. 
Then, the corresponding array response vectors of each in some azimuth-elevation $(\theta,\phi)$ relative to broadside are denoted by $\vatx(\theta,\phi) \in \setvectorcomplex{\Nt}$ and $\varx(\theta,\phi) \in \setvectorcomplex{\Nr}$.
We assume the user terminal is oriented such that the arrays' broadside directions are aligned with zenith and will reference all azimuth and elevations to the local coordinate system of the user terminal, as shown in \figref{fig:system}.

Let us assume the user terminal is interested in transmitting and receiving with the uplink and downlink satellites over a time horizon $\sT = \braces{t: t_1 \leq t \leq t_2}$.
In \leo systems, satellites are typically overhead a particular user for a few minutes at most, implying $\sT$ may occupy around two minutes or so.
At time $t$, let $(\thetadl_t,\phidl_t)$ be the direction from the user terminal to the downlink satellite and $(\thetaul_t,\phiul_t)$ be that toward the uplink satellite. 
Let $\powerdltx$ be the transmit power of the downlink satellite and $\gaindltx$ be the gain of its transmit antenna, assumed to be constant toward the user.
Let $\powerultx$ be the transmit power of the user terminal and $\powerdlnoise$ be its additive noise power.
Let $\gainulrx$ be the receive gain of the uplink satellite toward the user and $\powerulnoise$ be its additive noise power.
Let $\vhult \in \setvectorcomplex{\Nt}$ be the uplink channel from the user terminal's transmit array to the uplink satellite and  $\vhdlt \in \setvectorcomplex{\Nr}$ be the downlink channel from the downlink satellite to the user terminal's receive array, both at time $t$.
These channels are almost purely line-of-sight the majority of the time in practical LEO deployments.

This work will exclusively focus on beamforming at the user terminal, which we assume is accomplished using an analog beamforming architecture, following user terminals in practice today.
Let $\vf_t \in \setvectorcomplex{\Nt}$ be the beamforming weight vector employed by the user terminal to transmit to the uplink satellite at time $t$ and $\vw_t \in \setvectorcomplex{\Nr}$ be those used to receive from the downlink satellite at time $t$.
As a consequence of analog beamforming, the transmit and receive beams are restricted to some sets of physically realizable beams---i.e., $\vf_t \in  \sF$  and $\vw_t \in \sW$---capturing relevant power constraints, limited phase shifter resolution, and other hardware constraints.

Putting all this together, the uplink and downlink \glspl{snr} at time $t$ are correspondingly
\begin{align}
    \snrul_t 
    &= \frac{\powerultx \cdot \gainulrx \cdot \bars{\vf_t\ctrans \vhult}^2}{\powerulnoise}, \label{eq:snr-ul} \\
    \snrdl_t 
    &= \frac{\powerdltx \cdot \gaindltx \cdot \bars{\vw_t\ctrans \vhdlt}^2}{\powerdlnoise} \label{eq:snr-dl}. 
\end{align}
By operating in an in-band full-duplex fashion, the user terminal inflicts upon itself interference, the severity of which depends on the coupling of the transmit beam $\vf_t$ and receive beam $\vw_t$ across the \gls{mimo} self-interference channel $\mH \in \setmatrixcomplex{\Nr}{\Nt}$.
The \gls{inr} of this coupled self-interference is
\begin{align}
    \inrdl_t 
    &= \frac{\powerultx \cdot \bars{\vw_t\ctrans \mH \vf_t}^2}{\powerdlnoise}, \label{eq:inr-dl}
\end{align}
and the downlink \gls{sinr} at time $t$ is then
\begin{align}
    \sinrdl_t
    &= \frac{\snrdl_t}{1 + \inrdl_t}.
\end{align}
While \gls{snr} solely determines uplink performance, that on the downlink depends on the downlink \gls{sinr} and thus on both $\vf_t$ and $\vw_t$.
Treating self-interference as noise, the upper bound on the achievable sum spectral efficiency, for a given transmit beam $\vf_t$ and receive beam $\vw_t$, can be expressed as
\begin{align}
R_t(\vf_t,\vw_t) 
&= \logtwo{1+\snrul_t} + \logtwo{1+\sinrdl_t}. \label{eq:sum_se}
\end{align}

\section{Full-Duplex Beam Tracking} \label{sec:problem-formulation}

With the goal of enabling full-duplex operation at the user terminal, we aim to design its transmit and receive beams throughout the satellites' trajectories. 
In essence, this amounts to designing $(\vf_t,\vw_t)$ to maximize $\snrul_t$ and $\sinrdl_t$ for all $t \in \sT$ in pursuit of maximizing the sum rate. 
Formulating this as a proper optimization problem, we have
\begin{subequations} \label{eq:opt-1}
\begin{align}
\max_{\braces{(\vf_t,\vw_t)}_{t\in\sT}} \ &  \sum_{t\in\sT} R_t(\vf_t,\vw_t) \\
\subto \  
& \vf_t \in \sF, \ \vw_t \in \sW  \ \forall \ t \in \sT.
\end{align}
\end{subequations}
It is important to recognize that, in conventional LEO systems, it is reasonable to assume ground users have knowledge of satellite trajectories \textit{a priori}, given satellites follow carefully designed orbitals.
Applying conventional beam tracking to this problem would therefore amount to the user terminal steering its transmit beam directly toward the uplink satellite and its receive beam directly toward the downlink satellite.
While this would maximize the uplink and downlink \glspl{snr}, recent measurements have shown that this leads to prohibitively high self-interference (i.e., $\inrdl_t$) with high probability \cite{roberts_att_angular}. 
The same measurement campaign \cite{roberts_att_angular}, however, also showed that slightly shifting the transmit and receive beams in just the right way can substantially reduce self-interference.
We aim to leverage this phenomenon in constructing the beam tracking solution that follows.

Before unveiling our solution, it is important to first underscore noteworthy considerations when approaching problem \eqref{eq:opt-1}.
Since the user terminal has knowledge of the satellite trajectories \textit{a priori}  and the uplink and downlink channels are almost purely line-of-sight, it is reasonable to assume that the user is capable of calculating the uplink and downlink \glspl{snr} at any time $t$ for any given transmit beam $\vf_t$ and receive beam $\vw_t$.
While $\snrul_t$ and $\snrdl_t$ can be reliably calculated for a given beam pair $(\vf_t,\vw_t)$, it is less likely for a system to precisely know the self-interference $\inrdl_t$ coupled by these beams without explicitly measuring it.
This is for two main reasons.
First, as highlighted in the introduction, obtaining knowledge of the self-interference channel $\mH$ is not straightforward in practice due to its dimensionality and hardware uncertainties.
Second, even if one were to estimate $\mH$ perfectly, the actual self-interference coupled by a given beam pair can be difficult to predict exactly since these beams do not get realized perfectly in hardware, due to array miscalibration or phase shifter quantization and errors.
While these impairments may be negligible in many applications, full-duplex is sensitive to such since small changes in self-interference can degrade received SINR substantially. 
For these reasons, we take a measurement-driven approach to tackling this problem, as it appears to be the most reliable, given current understandings of self-interference in full-duplex systems operating at and around \mmwave frequencies.

\begin{figure}[t]
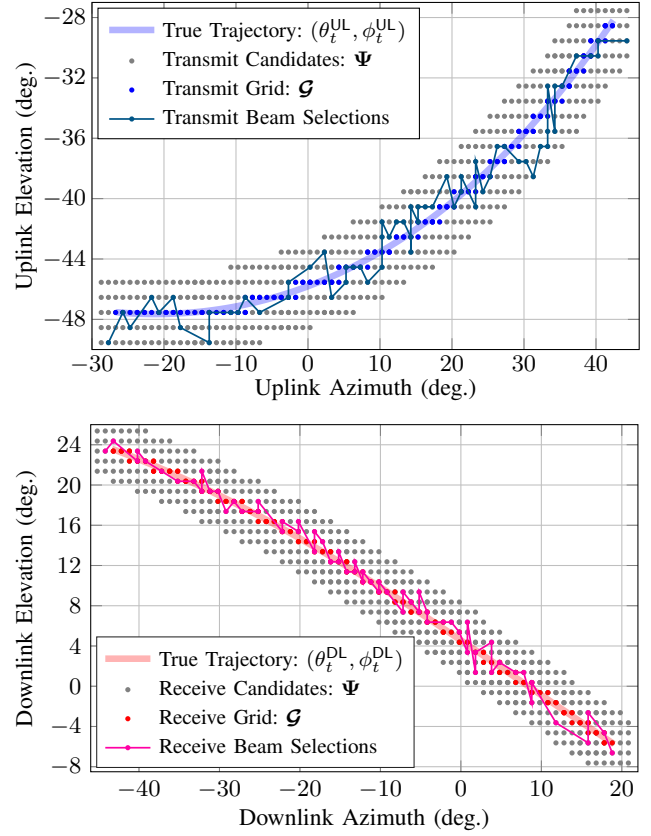

    \centering
    \vspace{0.01in}%
    
    \input{plots/traj_tx.tex}\\

    \vspace{0.25cm}
    
    \input{plots/traj_rx.tex}
    \caption{Visualizing the grid $\bm\sG$ and set of beam candidates $\bm\Psi$ constructed around the true trajectories of example uplink and downlink satellites for $\Delta\theta = \Delta\phi = 2^\circ$. All dots are spaced $1^\circ$ apart in azimuth and elevation. The beams eventually selected by our proposed scheme are overlaid as lines.}
    \label{fig:traj}
\end{figure}

\subsection{Grid Construction}
The first step in our design involves constructing a grid of transmit-receive steering directions, which we will later build upon to form a set of candidate transmit-receive beam pairs.
Recall, $\thphult$ and $\thphdlt$ denote the azimuth and elevation directly toward the uplink and downlink satellites at time $t$, respectively.
Let us then define the uplink-downlink direction tuple (in degrees) at time $t$ as
\begin{align}
\mOmega_t \triangleq \parens{\thetaul_{t},\phiul_{t},\thetadl_{t},\phidl_{t}}.
\end{align}
Recent measurements have indicated that slightly shifting the steering direction of the transmit and/or receive beams of a full-duplex \mmwave transceiver on the order of $1^\circ$ in azimuth and/or elevation can substantially reduce self-interference \cite{roberts_att_angular}. 
Inspired by this, we aim to construct a grid of beam candidates with $1^\circ$ resolution in azimuth and elevation, though our general approach could accommodate sub-$1^\circ$ resolutions as well.
Rather than merely quantizing $\mOmega_t$ to the nearest integer tuple at each time $t$, we instead aim to quantize it to a $1^\circ$-grid which is \textit{best fit} to the entire trajectory $\sT$.
Finding this best-fit $1^\circ$-grid amounts to shifting the integer $1^\circ$-grid by some bias in each of the four dimensions: uplink azimuth and elevation and downlink azimuth and elevation.
The appropriate bias applied to the uplink azimuth dimension, denoted by $\beta_\theta^\labelul$, can be found through exhaustive search by solving
\begin{align} \label{eq:beta-ul}
\beta_\theta^\labelul = \argmin_{\beta : |\beta| \leq 0.5} \ \sum_{t\in\sT} \parens{\thetaul_t + \beta - \nint{\thetaul_t + \beta}}^2,
\end{align}
where $\nint{a}$ rounds $a$ to the nearest integer.
Shifting the four-dimensional (4D) $1^\circ$-grid by $\beta_\theta^\labelul$ will then minimize the average distance from $\thetaul_t$ to its nearest grid point along the trajectory $t\in\sT$.
Calculating the other three bias factors analogously allows us to populate the bias vector 
\begin{align} \label{eq:beta}
\vbeta \triangleq \parens{\beta^\labelul_\theta,\beta^\labelul_\phi,\beta^\labeldl_\theta,\beta^\labeldl_\phi}, 
\end{align}
which allows us to construct the best-fit 4D $1^\circ$-grid as
\begin{align} \label{eq:grid}
\bm\sG = \bigcup_{t\in\sT} \nint{\mOmega_t + \vbeta } - \vbeta.
\end{align}
The grid $\bm\sG$ can equivalently be thought of as the \textit{quantized} trajectory of the uplink and downlink satellites, as seen in \figref{fig:traj}.
Note that if $\vbeta$ happens to be $\bm 0$, then the grid $\bm\sG$ will be comprised of uplink-downlink direction pairs with integer azimuths and elevations.
When $\bm\beta \neq \bm0$, the grid $\bm\sG$ will be a grid with $1^\circ$ spacing shifted off the integer grid.
By constructing $\bm\sG$ in this way, the candidate beam pairs that we will populate shortly will steer more closely toward the uplink and downlink satellites throughout their trajectories, allowing us to deliver higher \gpsnr while reducing self-interference.

\subsection{Beam Selection}
With $\bm\sG$ constructed, we now turn our attention to placing so-called \textit{spatial neighborhoods} around the points in $\bm\sG$ to populate candidate transmit-receive beam pairs, and then we will select beam pairs from this candidate set to maximize sum spectral efficiency, an approach similar to \cite{roberts_steer}.
To formalize a spatial neighborhood, let us define $\Delta\theta$ and $\Delta\phi$ (in degrees) as the maximum amount we will allow a transmit and receive beam to shift in azimuth and elevation, respectively.
For example, with $\Delta\theta=\Delta\phi=2^\circ$, the transmit beam and receive beam may each shift by at most $2^\circ$ in azimuth and elevation from a point on the $1^\circ$-grid $\bm\sG$.
With $\Delta\theta$ and $\Delta\phi$ denoting the size of the spatial neighborhood, let us assume the neighborhoods are constructed with $1^\circ$ resolution.
Put simply, let the azimuthal and elevational neighborhoods be defined as
\begin{align} 
\sN_\theta &= \braces{-\Delta\theta,\dots,-1,0,1,\dots,\Delta\theta} \label{eq:nbr-az} \\
\sN_\phi &= \braces{-\Delta\phi,\dots,-1,0,1,\dots,\Delta\phi}. \label{eq:nbr-el}
\end{align}
Then, a 2D neighborhood (in azimuth and elevation) with size $(\Delta\theta,\Delta\phi)$ and a resolution of $1^\circ$ can be constructed by the Cartesian product $\sN_\theta \times \sN_\phi$.
The 4D transmit-receive neighborhood would then correspondingly be
\begin{align} \label{eq:nbr}
\bm\sN = \sN_\theta \times \sN_\phi \times \sN_\theta \times \sN_\phi.
\end{align}
The set $\bm\sN$ contains $(2\Delta\theta+1)^2 \cdot (2\Delta\phi+1)^2$ different relative shifts a transmit-receive beam pair can apply.
Centering this neighborhood around each of the grid points in $\bm\sG$ amounts to
\begin{align} \label{eq:cand}
\bm\Psi = \bm\sG + \bm\sN = \braces{\mOmega + \bm\eta : \mOmega \in \bm\sG, \bm\eta \in \bm{\sN}},
\end{align}
containing candidate transmit-receive steering directions centered on the initial $1^\circ$-grid $\bm\sG$.
The set of transmit directions and receive directions in $\bm\Psi$ can be visualized in \figref{fig:traj}.
Note that, since $\bm\sG$ has $1^\circ$ resolution and $\bm\sN$ has $1^\circ$ resolution, their sum also has a resolution of $1^\circ$.
This is by design for two reasons.
First, as stated, it was found in \cite{roberts_att_angular} that shifting beams on the order of $1^\circ$ can substantially reduce self-interference.
Second, since both $\bm\sG$ and $\bm\sN$ have $1^\circ$ resolution, the cardinality of their sum will often be much less than the sum of their cardinalities, i.e.,  $\card{\bm\Psi} \ll \card{\bm\sG} + \card{\bm\sN}$.
This will save on measurements when implementing our proposed technique, as we will see.

Suppose the user terminal can form a transmit beam toward some $(\varthetaul,\varphiul)$ and a receive beam toward some $(\varthetadl,\varphidl)$ via so-called matched filter beamforming weights $\vf(\varthetaul,\varphiul) = \vatx(\varthetaul,\varphiul)$ and $\vw(\varthetadl,\varphidl) = \varx(\varthetadl,\varphidl)$.
Then, the self-interference when steering toward these directions corresponds to the form shown in \eqref{eq:inr-dl} with $\vf(\varthetaul,\varphiul)$ and $\vw(\varthetadl,\varphidl)$.
After forming the set of beam candidates $\bm\Psi$, we assume the user terminal collects measurements (or references past measurements) of self-interference for each beam candidate $\parens{\varthetaul,\varphiul,\varthetadl,\varphidl} \in \bm\Psi$.
With these measurements of self-interference and the assumption that $\snrul_t$ and $\snrdl_t$ can be computed based on the known satellite trajectories, the transmit and receive beams which maximize the sum spectral efficiency at any given time $t \in \sT$ can be found directly by solving the following.
\begin{subequations} \label{eq:opt-2}
\begin{align}
\max_{\parens{\varthetaul_t,\varphiul_t},\parens{\varthetadl_t,\varphidl_t}} \
& R_t\parens{\vf\parens{\varthetaul_t,\varphiul_t},\vw\parens{\varthetadl_t,\varphidl_t}} \\
\subto \  
&\parens{\varthetaul_t,\varphiul_t,\varthetadl_t,\varphidl_t} \in \bm\Psi
\end{align}
\end{subequations}
Notice that solving problem \eqref{eq:opt-2} amounts to a mere exhaustive search through $\bm\Psi$, which incurs trivial complexity.
This is a key advantage of the proposed technique over other search-based techniques. 
While the search space of $(\vf,\vw)$ is $\Nt\Nr$-dimensional (complex) for a single time instant, our approach reduces the search space to merely $\bm\Psi$.  
Given the small-scale variability of self-interference observed in \cite{roberts_steer,roberts_att_angular}, this dimensionality reduction has proven to reliably reduce self-interference without prohibitively degrading \gls{snr}, which we further confirm in the next section through simulation.
This concludes our proposed full-duplex beam tracking scheme, which is summarized concisely in \algref{alg:full}.

\begin{algorithm}[t]
    \caption{Proposed full-duplex beam tracking scheme.}%
    \label{alg:full}%
    \begin{algorithmic}
        \REQUIRE
        Trajectories $\braces{\mOmega_t : t \in \sT}$; neighborhood $(\Delta\theta,\Delta\phi)$ \\
        \STATE Calculate $\bm\beta$ using \eqref{eq:beta-ul} and populate the grid $\bm\sG$ using \eqref{eq:grid}.
        \STATE Form the neighborhood $\bm\sN = \sN_\theta \times \sN_\phi \times \sN_\theta \times \sN_\phi$.
        \STATE Construct candidates $\bm\Psi = \bm\sG + \bm\sN$. 
        \STATE For each $\parens{\varthetaul,\varphiul,\varthetadl,\varphidl} \in \bm\Psi$, measure $\inrdl$ at the user terminal with $\vf(\varthetaul,\varphiul)$ and $\vw(\varthetadl,\varphidl)$. 
        \FOR{$t \in \sT$} 
        \STATE Initialize: $R_t \leftarrow 0$.
        \FOR{$\parens{\varthetaul,\varphiul,\varthetadl,\varphidl} \in \bm\Psi$}
        \STATE Calculate $\snrul_t$ with $\vf(\varthetaul,\varphiul)$. 
        \STATE Calculate $\snrdl_t$ with $\vw(\varthetadl,\varphidl)$. 
        \STATE $\sinrdl_t \leftarrow \frac{\snrdl_t}{1+\inrdl}$ with $\vf(\varthetaul,\varphiul)$, $\vw(\varthetadl,\varphidl)$
        \STATE $R = \logtwo{1+\snrul_t} + \logtwo{1+\sinrdl_t}$
        \IF{$R > R_t$}
        \STATE $R_t \leftarrow R$; $\vf_t \leftarrow \vf\parens{\varthetaul,\varphiul}$; $\vw_t \leftarrow \vw(\varthetadl,\varphidl)$
        \ENDIF
        \ENDFOR
        \ENDFOR 
        \ENSURE{Transmit and receive beams $\braces{\parens{\vf_t,\vw_t}: t\in\sT}$}
    \end{algorithmic}
    \vspace{-0.1cm}
\end{algorithm}

\section{Simulation Results} \label{sec:results}
In this section, we assess our proposed full-duplex beam tracking scheme in terms of self-interference $\inrdl_t$, downlink \gls{sinr}, and uplink \gls{snr}, comparing each to what would be achieved nominally with a conventional beam tracking scheme directly following the satellites.
To accomplish this, we considered a user located on the UCLA campus at a longitude-latitude of $(34.0722^\circ, -118.4441^\circ)$, and simulated a LEO satellite system based on orbital parameters in public FCC filings for Amazon's emerging Project Kuiper constellation \cite{kuiper}.  
In total, the constellation was comprised of 3236 satellites operating at 20 GHz across 100 MHz at altitudes of 590--630 km.  
To select a downlink and uplink satellite pair at a given instant, we randomly selected two from the set of satellites which were overhead the user---above a minimum elevation angle of $35^\circ$---for at least two minutes.
Thus, the time horizon $\sT$ was two minutes in duration.

Satellite parameters were based on Ka-band 600 km altitude parameters from Table 6.1.1.1-2 in \cite{38.821}, and user terminal parameters were based on VSAT parameters from Table 6.1.1.1-3 in \cite{38.821}.
The transmit power of each satellite was set to 15.5 dBm, and that of the ground user terminal was 36 dBm. 
Each satellite was equipped with a 30.5 dBi antenna, and the user with a 16$\times$16 antenna array with a max gain of 29 dBi for transmitting and 39.7 dBi for receiving.  
The noise power of satellites was set to $-93.1$~dBm and that of the user was $-95.64$~dBm.
Self-interference was simulated using the model proposed in \cite{roberts_att_modeling}, which itself was based on nearly 20 million measurements taken with 16$\times$16 \mmwave phased arrays.

\begin{figure}[!t]
    \centering
    \subfloat[Self-interference, $\inrdl_t$.]{\input{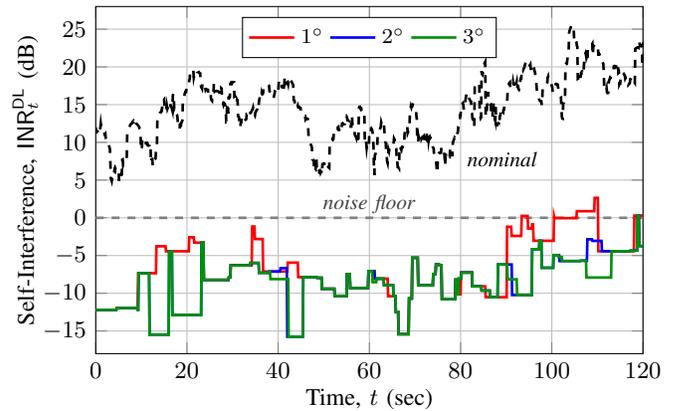}\label{fig:inr-traj}}
    \quad
    \subfloat[Uplink SNR and downlink SINR.]{\input{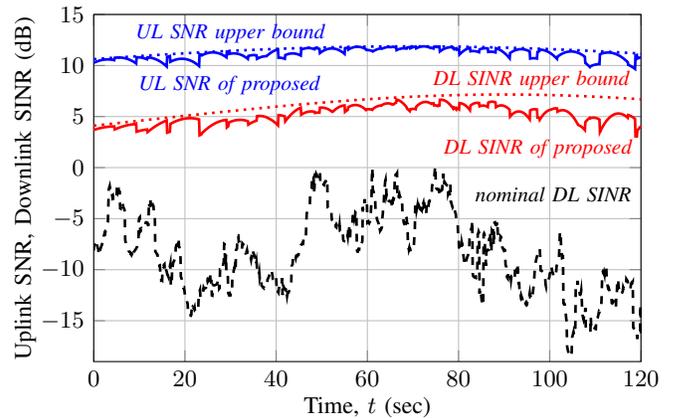}\label{fig:sinr-traj}}
    \caption{Self-interference, uplink SINR, and downlink SINR throughout the two-minute overpass of a downlink satellite and uplink satellite. The proposed scheme reliably reduces self-interference to below the noise floor, while attaining near-maximal downlink SINR and near-maximal uplink SNR.}
    \label{fig:inr-sinr-traj}
    \vspace{-0.4cm}%
\end{figure}

To begin, \figref{fig:inr-traj} depicts self-interference $\inrdl_t$ throughout the two-minute pass of a single pair of downlink and uplink satellites.
Shown is that with our proposed scheme for neighborhood sizes $\Delta\theta = \Delta\phi  = 1^\circ, 2^\circ, 3^\circ$ versus that with a conventional beam tracking scheme.
First, note that $\inrdl_t = 0$ dB corresponds to self-interference as strong as noise, below which is preferred to reap the benefits of full-duplex operation.
Naively steering directly toward the satellites, as done conventionally to maximize $\snrul_t$ and $\snrdl_t$, leads to self-interference which is typically more than 10 dB above the noise floor---too high for meaningful full-duplex operation.
With the proposed beam tracking solution, however, self-interference is reliably suppressed by around 20~dB to below the noise floor.
With $\Delta\theta = \Delta\phi = 1^\circ$, there are occasional instances where $\inrdl_t$ exceeds noise (about 8.2 seconds in total).
Widening the neighborhood to $2^\circ$ eliminates virtually all of these instances, and increasing further to $3^\circ$ offers occasional improvements beyond this. 

Now, in \figref{fig:sinr-traj}, we show the uplink \gls{snr} and downlink \gls{sinr} corresponding to the trajectory shown previously in \figref{fig:inr-traj}.
Conventional beam tracking maximizes uplink \gls{snr} (dotted blue) and downlink \gls{snr} (dotted red) but achieves poor downlink \gls{sinr}, due to the high $\inrdl_t$ seen in \figref{fig:inr-traj}.
The proposed scheme (with $\Delta\theta=\Delta\phi=2^\circ$) attains near-maximal uplink \gls{snr}, often falling short by only a fraction of a dB. 
Note that the downlink \gsnr attained with conventional beam tracking serves as the upper bound on downlink \gsinr.
The downlink \gsinr with conventional beam tracking falls well short of this upper bound since it ignores self-interference.
The downlink \gsinr with our proposed scheme, on the other hand, closely follows the upper bound, sacrificing at most around $3$ dB.
These results confirm that the proposed scheme is indeed capable of substantially reducing self-interference while delivering near-maximal \gls{snr} on both the downlink and uplink via slight, strategic shifts of the transmit and receive beams as the satellites traverse across the sky.

\begin{figure}[!t]
    \centering
    \subfloat[Self-interference, $\inrdl_t$.]{\input{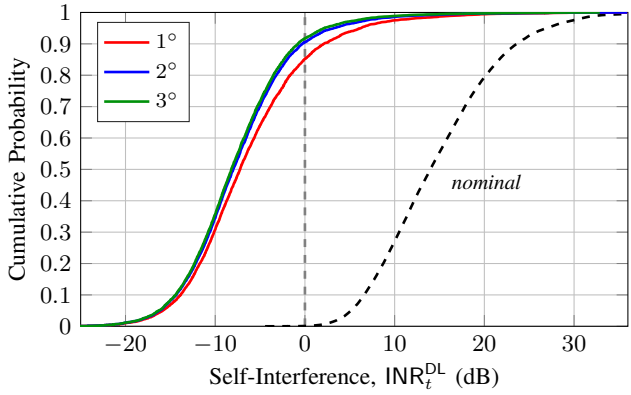}\label{fig:inr-cdf}}
    \quad
    \subfloat[Downlink SNR and downlink SINR.]{\input{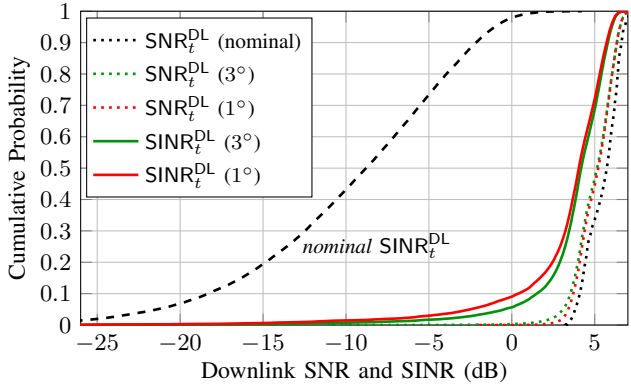}\label{fig:sinr-cdf}}
    \caption{Empirical CDFs of self-interference, downlink SNR, and downlink SINR across two-minute overpasses of 136 downlink-uplink satellite pairs.}
    \label{fig:inr-sinr-cdf}
    \vspace{-0.4cm}%
\end{figure}

To more broadly assess our scheme, we measured its performance across two-minute passes of 136 different downlink-uplink satellite pairs and examined the distribution of both $\inrdl_t$ and $\sinrdl_t$.
\figref{fig:inr-cdf} depicts the resulting empirical \gpcdf of self-interference with our proposed scheme and with conventional beam tracking.
Self-interference is almost always well above noise with conventional beam tracking.
Our proposed scheme, on the other hand, reliably reduces self-interference to below the noise floor 85\% of the time with a $1^\circ$ neighborhood and over 90\% of the time with $2^\circ$ or greater.
In median, our proposed scheme reduces $\inrdl_t$ by over 20 dB.
However, it is worth noting that there exist select conditions where slightly shifting the transmit and receive beams proves incapable of reducing self-interference to near or below the noise floor.
Perhaps resorting to half-duplex would be necessary in such cases.

Finally, in \figref{fig:sinr-cdf}, we plot the empirical \gpcdf of downlink \gsnr and \gsinr corresponding to \figref{fig:inr-cdf}.
The black dotted line is the \gcdf of $\snrdl_t$ attained by conventional beam tracking and is thus maximized, setting the upper bound on $\sinrdl_t$.
As seen thus far, the \gsinr with conventional beam tracking falls well short of this upper bound, since it is plagued by prohibitively high self-interference.
In stark contrast, the \gsinr with our proposed scheme typically falls short by only about 2 dB from the upper bound, due to residual $\inrdl_t$ and small sacrifices made in $\snrdl_t$ to reduce $\inrdl_t$. 
Our proposed scheme does exhibit a non-negligible lower tail, wherein the downlink \gsinr is less than $0$ dB about 10\% of the time with a $1^\circ$ neighborhood. 
This reduces to about 5\% of the time if the neighborhood widens to $3^\circ$.
Overall, these distributions confirm that our proposed full-duplex beam tracking scheme can reliably offer appreciable downlink \gpsinr by reducing self-interference to below the noise floor the vast majority of the time, regardless of which satellites are selected.

\section{Conclusion} \label{sec:conclusion}

This paper introduces a novel beam tracking scheme for full-duplex ground user terminals seeking to transmit uplink to one satellite while receiving downlink from another satellite at the same frequency.
Rather than directly follow the known trajectories of the two satellites, our scheme steers the transmit and receive beams of the user terminal strategically to reduce self-interference without compromising high \gsnr.
This is accomplished by leveraging a recent measurement campaign of self-interference which revealed that slight shifts of the transmit and receive beams can reliably reduce self-interference in full-duplex \mmwave phased array systems.
Through simulation, we showed that our proposed scheme can reduce self-interference to below the noise floor 85\%--90\% of the time, and this yields downlink SINRs which typically fall only about 2 dB short of their upper bound.
Valuable future work may include exploring other applications of full-duplex in LEO satellite systems, such as at the satellite or ground station, and the enhancements it may offer to such networks.

\bibliographystyle{bibtex/IEEEtran}
\bibliography{bibtex/IEEEabrv,refs}

\begin{thebibliography}{10}
\providecommand{\url}[1]{#1}
\csname url@samestyle\endcsname
\providecommand{\newblock}{\relax}
\providecommand{\bibinfo}[2]{#2}
\providecommand{\BIBentrySTDinterwordspacing}{\spaceskip=0pt\relax}
\providecommand{\BIBentryALTinterwordstretchfactor}{4}
\providecommand{\BIBentryALTinterwordspacing}{\spaceskip=\fontdimen2\font plus
\BIBentryALTinterwordstretchfactor\fontdimen3\font minus
  \fontdimen4\font\relax}
\providecommand{\BIBforeignlanguage}[2]{{%
\expandafter\ifx\csname l@#1\endcsname\relax
\typeout{** WARNING: IEEEtran.bst: No hyphenation pattern has been}%
\typeout{** loaded for the language `#1'. Using the pattern for}%
\typeout{** the default language instead.}%
\else
\language=\csname l@#1\endcsname
\fi
#2}}
\providecommand{\BIBdecl}{\relax}
\BIBdecl

\bibitem{IMT_framework}
ITU-R, ``Framework and overall objectives of the future development of {IMT}
  for 2030 and beyond,'' \emph{Report ITU-R M.2160-0}, Nov 2023.

\bibitem{kodheli_satellite}
O.~Kodheli \emph{et~al.}, ``Satellite communications in the new space era: A
  survey and future challenges,'' \emph{IEEE Commun. Surveys \& Tutor.},
  vol.~23, no.~1, pp. 70--109, 2021.

\bibitem{smida_fd_phy_2024}
B.~Smida \emph{et~al.}, ``In-band full-duplex: The physical layer,''
  \emph{Proc. {IEEE}}, vol. 112, no.~5, pp. 433--462, May 2024.

\bibitem{baeza_overview_2023}
V.~M. Baeza \emph{et~al.}, ``Overview of use cases in single channel full
  duplex techniques for satellite communication,'' in \emph{Intl. Commun.
  Satellite Sys. Conf.}, vol. 2023, 2023, pp. 250--254.

\bibitem{Flexible}
\emph{{Discussion on supporting TDD duplex scheme for NTN}}, document
  R1-1912538, TSG RAN WG1 \#99, Reno, USA, Nov. 2019.

\bibitem{roberts_wcm}
I.~P. Roberts, J.~G. Andrews, H.~B. Jain, and S.~Vishwanath, ``Millimeter-wave
  full duplex radios: New challenges and techniques,'' \emph{{IEEE} Wireless
  Commun.}, vol.~28, no.~1, pp. 36--43, Feb. 2021.

\bibitem{choi_circular_2024}
D.~Choi and G.~Byun, ``Circular polarization conversion using dual frequency
  selective surfaces for full-duplex satellite communications,'' \emph{IEEE
  Access}, vol.~12, pp. 120\,219--120\,225, 2024.

\bibitem{erkelenz_hybrid_2024}
K.~Erkelenz, N.~Sielck, A.~Koelpin, and A.~F. Jacob, ``A hybrid-integrated
  {K}-/{Ka}-band phased array module with dual-polarized shared aperture,''
  \emph{{IEEE} Trans. Microw. Theory Techn.}, pp. 1--10, 2024.

\bibitem{sazegar_kuband_2023}
M.~Sazegar \emph{et~al.}, ``Ku-band {SATCOM} user terminal with complete beam
  steering using a shared aperture metasurface for full-duplex operation,'' in
  \emph{European Conf. Antennas Propagation}, 2023, pp. 1--3.

\bibitem{navarro_ultra_2017}
J.~Navarro, ``Ultra-small aperture terminals for {SATCOM} on-the-move
  applications,'' in \emph{IEEE MTT-S Intl. Microw. Symp.}, 2017, pp.
  1152--1154.

\bibitem{abedian_mmwave_2023}
M.~Abedian \emph{et~al.}, ``{MM-Wave} high isolated dual polarized dielectric
  resonator antenna for in-band full-duplex systems,'' \emph{IEEE Access},
  vol.~11, pp. 38\,218--38\,225, 2023.

\bibitem{roberts_steer}
I.~P. {Roberts}, A.~Chopra, T.~Novlan, S.~Vishwanath, and J.~G. {Andrews},
  ``\textsc{Steer}: Beam selection for full-duplex millimeter wave
  communication systems,'' \emph{{IEEE} Trans. Commun.}, vol.~70, no.~10, pp.
  6902--6917, 2022.

\bibitem{roberts_att_angular}
I.~P. {Roberts}, A.~Chopra, T.~Novlan, S.~Vishwanath, and J.~G. {Andrews},
  ``Beamformed self-interference measurements at 28 {GHz}: Spatial insights and
  angular spread,'' \emph{{IEEE} Trans. Wireless Commun.}, vol.~21, no.~11, pp.
  9744--9760, Jun. 2022.

\bibitem{kuiper}
\emph{{Application of {Kuiper Systems LLC} for authority to launch and operate
  a non-geostationary satellite orbit system in Ka-band frequencies}}, Jul.
  2019, {Kuiper} Systems LLC.

\bibitem{38.821}
\emph{{Solutions for NR to support Non-Terrestrial Networks (NTN)}}, 2021, 3GPP
  \textrm{T}R 38.821 Version 16.1.0.

\bibitem{roberts_att_modeling}
I.~P. Roberts, A.~Chopra, T.~Novlan, S.~Vishwanath, and J.~G. Andrews,
  ``Spatial and statistical modeling of multi-panel millimeter wave
  self-interference,'' \emph{{IEEE} J. Sel. Areas Commun.}, vol.~41, no.~9, pp.
  2780--2795, Sep. 2023.

\end{thebibliography}

\end{document}